\documentclass[12pt]{article}

\RequirePackage[OT1]{fontenc}
\usepackage[english]{babel}
\usepackage{amsthm,amsmath,amssymb,amsfonts,xcolor,graphicx,natbib,color,url,enumerate}
\usepackage[plain,noend]{algorithm2e}
\RequirePackage[colorlinks,citecolor=blue,urlcolor=blue]{hyperref}
\def\frac#1#2{{\textstyle{#1\over#2}}}

\DeclareSymbolFont{AMSb}{U}{msb}{m}{n}
\DeclareMathSymbol{\Natural}{\mathbin}{AMSb}{"4E}
\DeclareMathSymbol{\Integer}{\mathbin}{AMSb}{"5A}
\DeclareMathSymbol{\Real}{\mathbin}{AMSb}{"52}
\DeclareMathSymbol{\Rational}{\mathbin}{AMSb}{"51}
\DeclareMathSymbol{\Imaginary}{\mathbin}{AMSb}{"49}
\DeclareMathSymbol{\Complex}{\mathbin}{AMSb}{"43} 
\DeclareMathSymbol{\Disk}{\mathbin}{AMSb}{"44} 
\def\bi{\begin{itemize}}
\def\ei{\end{itemize}}
\def\bd{\begin{description}}
\def\ed{\end{description}}
\def\ben{\begin{enumerate}}
\def\een{\end{enumerate}}
\def\bv{\begin{verbatim}}
\def\ev{\end{verbatim}}

\def\calP{{\mathcal{P}}}

\def\calS{{\mathcal{S}}}

\def\hat#1{{\widehat{#1}}}

\def\T{{ \mathrm{\scriptscriptstyle T} }}

\def\pr{{\rm Pr}}

\def\2to{{\ {\buildrel 2\over \longrightarrow}\ }}

\def\I1ton{{$I_1,\ldots,I_n$}}
\def\X1ton{{$X_1,\ldots,X_n$}}
\def\Y1ton{{$Y_1,\ldots,Y_n$}}
\def\Z1ton{{$Z_1,\ldots,Z_n$}}
\def\R1ton{{$R_1,\ldots,R_n$}}
\def\e1ton{{$e_1,\ldots,e_n$}}
\def\t1ton{{$t_1,\ldots,t_n$}}
\def\x1ton{{$x_1,\ldots,x_n$}}
\def\y1ton{{$y_1,\ldots,y_n$}}
\def\z1ton{{$z_1,\ldots,z_n$}}
\def\calP{{\mathcal{P}}}
\def\calS{{\mathcal{S}}}

\begin{document}
\thispagestyle{empty}
\baselineskip=28pt
\vskip 5mm
\begin{center} {\Large{\bf Full likelihood inference for max-stable data}}
\end{center}

\baselineskip=12pt

\vskip 5mm

\begin{center}
\large
Rapha\"el Huser$^1$, Cl\'ement Dombry$^2$, Mathieu Ribatet$^3$, Marc G. Genton$^1$
\end{center}

\footnotetext[1]{
\baselineskip=10pt CEMSE Division, King Abdullah University of Science and Technology (KAUST), \\Thuwal 23955-6900, Saudi Arabia. E-mail: raphael.huser@kaust.edu.sa and marc.genton@kaust.edu.sa}
\footnotetext[2]{
\baselineskip=10pt Department of Mathematics, University of Franche-Comt\'e, 25030 Besan\c con cedex, France. E-mail: clement.dombry@univ-fcomte.fr}
\footnotetext[3]{
\baselineskip=10pt Department of Mathematics, University of Montpellier, 34095 Montpellier cedex 5, France. E-mail: mathieu.ribatet@umontpellier.fr}

\baselineskip=17pt
\vskip 4mm
\centerline{\today}
\vskip 6mm

%%%%%%%%%%%%%%%%%%%%%%%%%%%%%%%%%%%%%%%%%%%%%%%%%%%%%%%%%%%%%%%%%%%%%%%%
\begin{center}
{\large{\bf Abstract}}
\end{center}

We show how to perform full likelihood inference for max-stable multivariate distributions or processes based on a stochastic Expectation-Maximisation algorithm, which combines statistical and computational efficiency in high-dimensions. The good performance of this methodology is demonstrated by simulation based on the popular logistic and Brown--Resnick models, and it is shown to provide dramatic computational time improvements with respect to a direct computation of the likelihood. Strategies to further reduce the computational burden are also discussed.
\baselineskip=16pt

\par\vfill\noindent

\pagenumbering{arabic}
\baselineskip=24pt

%%%%%%%%%%%%%%%%%%%%%%%%%%%%%%%%%
%%%%%%%%%%%%%%%%%%%%%%%%%%%%%%%%%
%%%%%%%%%%%%%%%%%%%%%%%%%%%%%%%%%

\newpage

\section{Introduction}
Under mild conditions, max-stable distributions and processes are useful for studying high-dimensional extreme events recorded in space and time \citep{Padoan.etal:2010,Davis.etal:2013b,Davison.etal:2013,deCarvalho.Davison:2014,Huser.Davison:2014a,Huser.Genton:2016}. This broad but constrained class of models may, at least theoretically, be used to extrapolate into the joint tail, hence providing a justified framework for risk assessment of extreme events. The probabilistic justification is that the max-stable property arises in limiting models for suitably renormalised maxima of independent and identically distributed processes; see, e.g., \citet{Davison.etal:2012}, \citet{Davison.Huser:2015} and \citet{Davison.etal:2018}. 

Because extremes are rare by definition, it is crucial for reliable estimation and prediction to extract as much information from the data as possible. Thus, efficient estimators play a particularly important role in statistics of extremes, and the maximum likelihood estimator is a natural choice thanks to its appealing large-sample properties. However, the likelihood function is excessively difficult to compute for high-dimensional data following a max-stable distribution. As detailed in \S\ref{inference}, likelihood evaluations require the computation of a sum indexed by all elements of a given set $\calP_D$, the cardinality of which grows more than exponentially with the dimension, $D$. In a thorough simulation study, \citet{Castruccio.etal:2016} stated that current technologies are limiting full likelihood inference to dimension $12$ or $13$, and they concluded that without meaningful methodological advances, a direct full likelihood approach will not be feasible.

To circumvent this computational bottleneck, several strategies have been advocated. \citet{Padoan.etal:2010} proposed a pairwise likelihood approach, combining the bivariate densities of carefully chosen pairs of observations. Although this method is computationally attractive and inherits many good properties from the maximum likelihood estimator, it also entails a loss in efficiency, which becomes more apparent in high dimensions \citep{Huser.etal:2016}. More efficient triplewise and higher-order composite likelihoods were investigated by \citet{Genton.etal:2011}, \citet{Huser.Davison:2013a}, \citet{Sang.Genton:2014} and \citet{Castruccio.etal:2016}. However, they are still not fully efficient, and it is not clear how to optimally select the composite likelihood terms. Furthermore, because composite likelihoods are generally not valid likelihoods \citep{Varin.etal:2011}, the classical likelihood theory cannot be blindly applied for uncertainty assessment, testing, model validation and selection, and so forth.

%it is also not as easy to assess uncertainty, perform model validation and selection, and embed them within a Markov chain Monte Carlo (MCMC) algorithm (see nevertheless \citealp{Ribatet.etal:2012}). 

Alternatively, \citet{Stephenson.Tawn:2005} suggested augmenting the componentwise block maxima data $z^n=(z_{1}^{n},\ldots,z_{D}^{n})^{\rm T}$, where $n$ is the block size, with their occurrence times. This extra information may be summarized by an observed partition $\pi^n$ of the set $\{1,\ldots,D\}$, which indicates whether or not these maxima occurred simultaneously. Essentially, the Stephenson--Tawn likelihood corresponds the limiting joint ``density'' of $z^n$ and $\pi^n$, as $n\to\infty$, and it yields drastic simplifications and improved efficiency. However, \citet{Wadsworth:2015} and \citet{Huser.etal:2016} noted that this approach may be severely biased for finite $n$, especially in low-dependence scenarios. By fixing the limit partition, $\pi$, to the observed one, $\pi^n$, a strong constraint is imposed, creating model misspecification, to which likelihood methods are sensitive. %A related approach is to take advantage of the limiting Poisson point process representation of extremes, yielding efficient inference methods based on a variety of threshold-based likelihoods (see \citealp{Huser.etal:2016}). In particular, the censored Poisson likelihood and Stephenson--Tawn likelihood coincide when the marginal thresholds are taken to be the observed maxima $z^n$ \citep{Wadsworth.Tawn:2014}. Thus, their efficiency and robustness properties are similar \citep{Huser.etal:2016}.

In this paper, to mitigate the sub-asymptotic bias due to fixing the partition, we suggest returning to the original likelihood formulation, which integrates out the partition rather than treating it as known. By interpreting the limit partition $\pi$ as missing data, we show how to design a stochastic Expectation-Maximisation algorithm \citep{Dempster.etal:1977,Nielsen:2000} for efficient inference. The quality of the stochastic approximation to the full likelihood can be controlled and set to any arbitrary precision at a computational cost. We show that higher-dimensional max-stable models may be fitted in reasonable time. Importantly, our method is based solely on max-stable data and does not require extra information about the partition or the original processes, unlike the Stephenson--Tawn or related threshold-based methods. Our approach is based on the algorithm of \citet{Dombry.etal:2013} for conditional simulation of the partition given the data, and it is closely related to the recent papers of \citet{Thibaud.etal:2016} and \citet{Dombry.etal:2017}, who in a Bayesian setting developed a Markov chain Monte Carlo algorithm for max-stable processes by treating the partition as a latent variable to be resampled at each iteration.

%The paper is organized as follows: in \S2, preliminaries on max-stable distributions and processes are recalled. \S3 details the full and Stephenson--Tawn likelihoods, and describes our novel stochastic Expectation-Maximisation algorithm to maximise the former. A simulation study is performed in \S4 based on the multivariate logistic model; estimation performance, computational burden, and strategies to speed up the algorithm are analyzed in this framework. \S5 concludes with a brief discussion.

\section{Max-stable processes and distributions}\label{maxstable}
\subsection{Definition, construction, and models}
Consider a sequence of independent and identically distributed processes $Y_1(s),Y_2(s),\ldots,$ indexed by spatial site $s\in\calS\subset\Real^d$, and assume that there exist sequences of functions $a_n(s)>0$ and $b_n(s)$, such that the renormalised pointwise block maximum process (with block size $n$)
\begin{equation}\label{maximum}
Z^n(s)=a_n(s)^{-1}\left[\max\{Y_1(s),\ldots,Y_n(s)\}-b_n(s)\right]
\end{equation}
converges in the sense of finite-distributional distributions, as $n\to\infty$, to a process $Z(s)$ with non-degenerate marginal distributions, i.e., $Y(s)$ is in the max-domain of attraction of $Z(s)$. Then, the limit $Z(s)$ is max-stable (see, e.g., \citealp{deHaan.Ferreira:2006}, Chap.~9). That is, pointwise maxima of independent copies of the limit process $Z(s)$ have the same dependence structure as $Z(s)$ itself, while marginal distributions are in the same location-scale family and coincide with the generalized extreme-value distribution.%; see \citet{Davison.etal:2013} for precise mathematical details.

% in the sense that when the processes $Y_1(s),\ldots,Y_n(s)$ in \eqref{maximum} are substituted by $n$ independent copies of $Z(s)$, then the pointwise block maximum process $Z^n(s)$ is \emph{equal} in distribution to $Z(s)$, for any integer $n$. By definition of a max-stable process, all finite-dimensional distributions are max-stable. In particular, univariate margins are generalised extreme-value distributed. 

\begin{figure}[t!]
\centering
\includegraphics[width=0.8\linewidth]{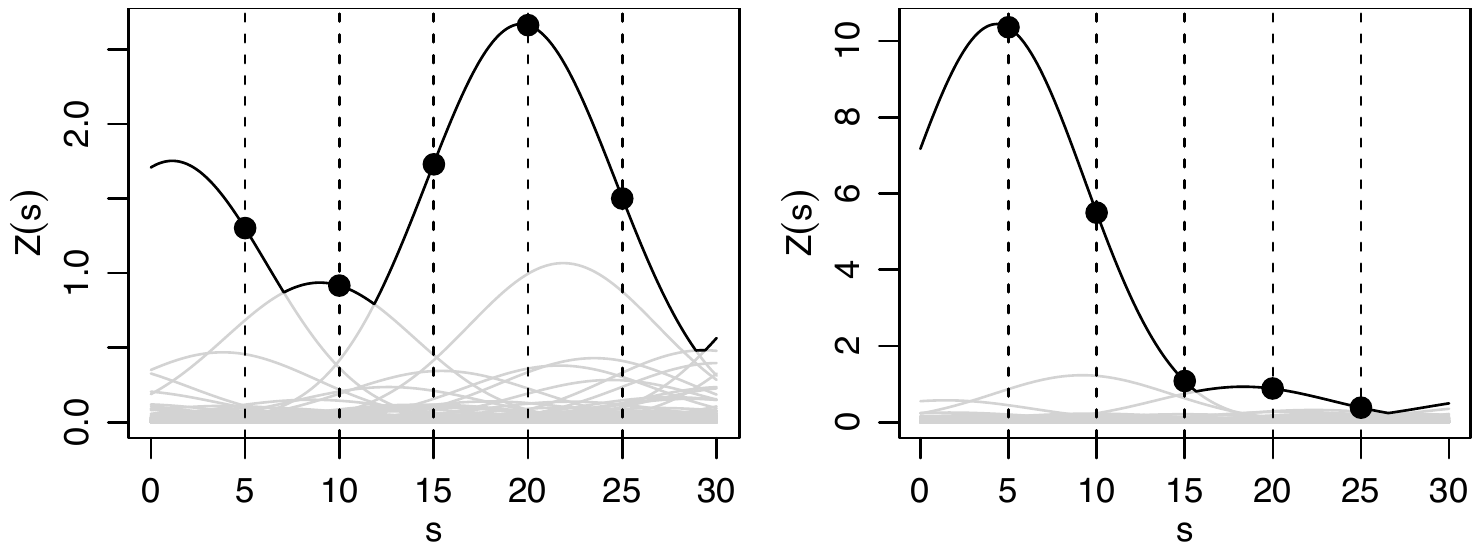}
\caption{Two realisations (black) from the same \citet{Smith:1990b} max-stable process on the line defined by setting $W(s)=\phi(s-U;\sigma^2)$, $s\in\calS=\Real$, in \eqref{representation}, with the latent profiles $W_j(s)/P_j$ (grey). Here, $\sigma^2=5$. When observed at sites $5,10,15,20,25$, the partitions are $\pi=\{\{1\},\{2\},\{3,4,5\}\}$ (left) and $\pi=\{\{1,2,3\},\{4,5\}\}$ (right).}\label{Fig:simul}
\end{figure}

Consider now points of a unit rate Poisson point process, $P_1,P_2,\ldots$, and independent copies, $W_1(s),W_2(s),\ldots$, of a stochastic process $W(s)\geq0$ with unit mean. Then, the process
\begin{equation}\label{representation}
Z(s)=\sup_{j\geq 1} W_j(s)/P_j,\quad s\in\calS,
\end{equation} 
is max-stable with unit Fr\'echet marginal distributions, i.e., $\pr\{Z(s)\leq z\}=\exp(-1/z)$, $z>0$ \citep{deHaan:1984,Schlather:2002}. In the remaining of the paper, we shall always consider max-stable processes $Z(s)$ with unit Fr\'echet marginal distributions. Representation \eqref{representation} is useful to build a wide variety of max-stable processes \citep{Smith:1990b,Schlather:2002,Kabluchko.etal:2009,Opitz:2013a,Xu.Genton:2016}, and to simulate from them \citep{Schlather:2002,Dombry.etal:2016}. Multivariate max-stable models can be constructed similarly. From \eqref{representation}, we deduce that the joint distribution at a finite set of sites $\calS_D=\{s_1,\ldots,s_D\}\subset\calS$ may be expressed as
\begin{equation}\label{distribution}
\pr\{Z(s_1)\leq z_1,\ldots,Z(s_D)\leq z_D\}=\exp\{-V(z_1,\ldots,z_D)\},
\end{equation} 
where the exponent function $V(z_1,\ldots,z_D)=E\left[\max\left\{{W(s_1)/z_1},\ldots,{W(s_D)/z_D}\right\}\right]$ satisfies homogeneity and marginal constraints \citep[see, e.g.,][]{Davison.Huser:2015}. As an illustration, Fig.~\ref{Fig:simul} shows two independent realisations from the same \citet{Smith:1990b} model defined by taking $W_j(s)=\phi(s-U_j;\sigma^2)$, $s\in\calS=\Real$, in \eqref{representation}, where $\phi(\cdot;\sigma^2)$ is the normal density with zero mean and variance $\sigma^2$, and the $U_j$s are points from a unit rate Poisson point process on the real line. 
%In \S\ref{simul}, we focus on the multivariate logistic model with exponent function $V(z_1,\ldots,z_D)=(\sum_{j=1}^D z_j^{-1/\theta})^\theta$, $\theta\in\Theta=(0,1]$, whilst providing further results in the supplementary material for the more realistic Brown--Resnick \citep{Kabluchko.etal:2009} process, which is defined by taking $W(s)=\exp\{\varepsilon(s)-\gamma(s)/2\}$, where $\varepsilon(s)$ is an intrinsically stationary Gaussian process with variogram $\gamma(h)={\rm var}\{\varepsilon(s)-\varepsilon(s+h)\}$, and such that $\varepsilon(0)=0$ almost surely. When $\gamma(h)=h^2/5$, we recover the Smith model displayed in Fig.~\ref{Fig:simul}.

\subsection{Underlying partition and extremal functions}
At each site $s\in\calS$, the pointwise supremum, $Z(s)$, in \eqref{representation} is realised by a single profile $W_j(s)/P_j$ almost surely. Such profiles are called extremal functions in \citet{Dombry.etal:2013}. As illustrated in Fig.~\ref{Fig:simul}, the extremal functions are only partially observed on $\calS_D=\{s_1,\ldots,s_D\}$; they define a random partition $\pi=\{\tau_1,\ldots,\tau_{|\pi|}\}$ (of size $|\pi|$) of the set $\{1,\ldots,D\}$, called hitting scenario in \citet{Dombry.etal:2013} that identifies clusters of variables stemming from the same event. For example, the partition $\pi=\{\{1\},\{2\},\{3,4,5\}\}$ on the left panel of Fig.~\ref{Fig:simul} indicates that the max-stable process at these five sites came from three separate independent events; in particular, the maxima at $s_3=15$, $s_4=20$, and $s_5=25$ were generated from the same profile.

Similarly, an observed partition $\pi^n$ of $\{1,\ldots,D\}$ may be defined for $Z^n(s)$ in \eqref{maximum} based on the original processes $Y_1(s),\ldots,Y_n(s)$. The knowledge of $\pi^n$ tells us if extreme events at different sites occurred simultaneously or not, so $\pi^n$ carries information about the strength of spatial extremal dependence. If the processes $Y_1(s),\ldots,Y_n(s)$ (suitably marginally transformed) are in the max-domain of attraction of $Z(s)$ in \eqref{representation}, then the partition $\pi^n$ converges in distribution to $\pi$, as $n\to\infty$, on the space of all partitions $\calP_D$ of $\{1,\ldots,D\}$ \citep{Stephenson.Tawn:2005}.

We now describe likelihood inference for max-stable vectors: by exploiting the information on the partition $\pi$ (Stephenson--Tawn likelihood), or by integrating it out (full likelihood).

\section{Likelihood inference}\label{inference}

\subsection{Full and Stephenson--Tawn likelihoods}
By differentiating the distribution \eqref{distribution} with respect to the variables $z_1,\ldots,z_D$, we can deduce that the corresponding density, or full likelihood for one replicate, may be expressed as
\begin{equation}\label{density}
g_{\rm Full}(z_1,\ldots,z_D)=\exp\{-V(z_1,\ldots,z_D)\}\sum_{\pi\in\calP_D}\prod_{i=1}^{|\pi|}\{-V_{\tau_i}(z_1,\ldots,z_D)\},
\end{equation} 
where $V_{\tau_i}$ denotes the partial derivative of the function $V$ with respect to the variables indexed by the set $\tau_i\subset\{1,\ldots,D\}$ \citep{Huser.etal:2016,Castruccio.etal:2016}. The sum in \eqref{density} is taken over all elements of $\calP_D$, the size of which equals the Bell number of order $D$, leading to an explosion of terms, even for moderate $D$. Each term in \eqref{density} corresponds to a different configuration of the profiles $W_j(s)/P_j$ in \eqref{representation} at the sites $s_1,\ldots,s_D$. Thus, \citet{Castruccio.etal:2016} argued that the computation of \eqref{density} is limited to dimension $12$ or $13$ with modern computational resources. 

As detailed in the Supplementary Material using a point process argument, and originally shown by \citet{Stephenson.Tawn:2005}, the joint density of the max-stable data $z=(z_1,\ldots,z_D)^{\rm T}$ and the associated partition $\pi=\{\tau_1,\ldots,\tau_{|\pi|}\}\in\calP_D$ is simply equal to
\begin{equation}\label{ST}
g_{\rm ST}(z_1,\ldots,z_D,\pi)=\exp\{-V(z_1,\ldots,z_D)\}\prod_{i=1}^{|\pi|}\{-V_{\tau_i}(z_1,\ldots,z_D)\},
\end{equation}
hence reducing the problematic sum to a single term, making likelihood inference possible in higher dimensions and simultaneously improving statistical efficiency. Because the asymptotic partition $\pi$ is not observed, \citet{Stephenson.Tawn:2005} suggested replacing it by the observed partition $\pi^n$ of occurrence times of maxima, which converges to $\pi$ provided the asymptotic model is well specified.  
%This ingenious idea was implemented, for example, by \citet{Davison.Gholamrezaee:2012} in a study of extreme temperatures in Switzerland. 
However, \citet{Wadsworth:2015} and \citet{Huser.etal:2016} showed that lack of convergence of $\pi^n$ to $\pi$ may result in severe estimation bias, which is especially strong in low-dependence cases. To circumvent this problem, \citet{Wadsworth:2015} proposed a bias-corrected likelihood; alternatively, we show in the next section how to design a stochastic Expectation-Maximisation algorithm to maximise \eqref{density}, while taking advantage of the computationally appealing nature of \eqref{ST}.

\subsection{Stochastic Expectation-Maximisation algorithm}
It is instructive to rewrite the full likelihood \eqref{density} using \eqref{ST} as $g_{\rm Full}(z_1,\ldots,z_D)=\sum_{\pi\in\calP_D}g_{\rm ST}(z_1,\ldots,z_D,\pi)$ because it highlights that the full likelihood simply integrates out the latent random partition $\pi$ needed for the Stephenson--Tawn likelihood. Interpreting $\pi$ as a missing observation and the Stephenson--Tawn likelihood as the completed likelihood, an Expectation-Maximisation algorithm \citep{Dempster.etal:1977} may be easily formulated. Assume that the exponent function $V(z_1,\ldots,z_D\mid \theta)$ is parametrised by a vector $\theta\in\Theta\subset\Real^p$. Starting from an initial guess $\theta_0\in\Theta$, the Expectation-Maximisation algorithm consists of iterating the following E- and M-steps for $r=1,\ldots,R$:
\begin{itemize}
\item[$\bullet$] E-step: compute the functional
\begin{equation}\label{Estep}
Q(\theta,\theta_{r-1})=E_{\pi\mid z,\theta_{r-1}}\left[\log \left\{g_{\rm ST}(z,\pi\mid \theta)\right\}\right]=\sum_{\pi\in\calP_D}g(\pi\mid z,\theta_{r-1})\log\{g_{\rm ST}(z,\pi\mid \theta)\},
\end{equation}
where the expectation is computed with respect to the discrete conditional distribution of $\pi$ given the data $z=(z_1,\ldots,z_D)^{\rm T}$ and the current value of the parameter $\theta_{r-1}$, i.e., 
 \begin{equation}\label{conditionaldistribution}
 g(\pi\mid z,\theta_{r-1})={g_{\rm ST}(z,\pi\mid \theta_{r-1})/ g_{\rm Full}(z\mid \theta_{r-1})}.
 \end{equation}
\item[$\bullet$] M-step: update the parameter as $\theta_{r}=\arg\max_{\theta\in\Theta}Q(\theta,\theta_{r-1})$.
\end{itemize}
\citet{Dempster.etal:1977} showed that the Expectation-Maximisation algorithm has appealing properties; in particular, the value of the log-likelihood increases at each iteration, which ensures convergence of $\theta_r$ to a local maximum, as $r\to\infty$. In our case, however, the expectation in \eqref{Estep} is tricky to compute: it contains again the sum over the set $\calP_D$, and \eqref{conditionaldistribution} relies on the full density $g_{\rm Full}(z\mid\theta_{r-1})$, which we try to avoid. To circumvent this issue, one solution is to approximate \eqref{Estep} by Monte Carlo as 
%\begin{itemize}
%\item[$\bullet$] Stochastic E-step: compute the random functional
\begin{equation}\label{StEstep}
\hat Q(\theta,\theta_{r-1})={1\over N}\sum_{i=1}^N \log \left\{g_{\rm ST}(z,\pi_i\mid \theta)\right\},\quad \pi_1,\ldots,\pi_N\sim  g(\pi\mid z,\theta_{r-1}),
\end{equation}
%\end{itemize}
where the partitions $\pi_1,\ldots,\pi_N$ are conditionally independent at best, or form an ergodic sequence at least. As $g(\pi\mid z,\theta_{r-1})\propto g_{\rm ST}(z,\pi\mid \theta_{r-1})$, see \eqref{conditionaldistribution}, it is possible to devise a Gibbs sampler to generate approximate simulations from $g(\pi\mid z,\theta_{r-1})$ without explicitly computing the constant factor $g_{\rm Full}(z\mid \theta_{r-1})$ in the denominator of \eqref{conditionaldistribution}. Thanks to ergodicity of the resulting Markov chain, the precision of the approximation \eqref{StEstep} may be set arbitrarily high by letting $N\to\infty$ (and discarding some burn-in iterations). More details about the practical implementation of the Gibbs sampler are given in \citet{Dombry.etal:2013} and in the Supplementary Material. Although the number of iterations of the Gibbs sampler, $N$, will typically be much smaller than the cardinality of $\calP_D$, the approximation \eqref{StEstep} to \eqref{Estep} will likely be reasonably good for moderate values of $N$ because only a few partitions $\pi\in\calP_D$ may be plausible or compatible with the data $z=(z_1,\ldots,z_D)^{\rm T}$. 

The asymptotic properties of the stochastic Expectation-Maximisation estimator, $\hat\theta_{\rm SEM}$, were studied in details by \citet{Nielsen:2000} and compared with the classical maximum likelihood estimator, $\hat\theta$; see \S2--3 therein, in particular Theorem 2. \citet{Dombry.etal:2017b} showed that the maximum likelihood estimator $\hat\theta$ is consistent and asymptotically normal for the most popular max-stable models, including the logistic and Brown--Resnick models used in this paper. This suggests that these appealing asymptotic properties should also be satisfied for the estimator $\hat\theta_{\rm SEM}$, provided some additional rather technical regularity conditions detailed in \citet{Nielsen:2000} are satisfied. If so, then the asymptotic performance of $\hat\theta_{\rm SEM}$ is akin to that of $\hat\theta$, though with a larger asymptotic variance. Finally, the inherent variability of the stochastic Expectation-Maximisation algorithm may also be a blessing: unlike the deterministic Expectation-Maximisation algorithm, it is less likely to get stuck at a local maximum of the full likelihood.

\section{Simulation study}\label{simul}
To assess the performance of the stochastic Expectation-Maximisation estimator $\hat\theta_{\rm SEM}$, we simulated data from the multivariate logistic max-stable distribution with exponent function $V(z_1,\ldots,z_D\mid\theta)=(\sum_{j=1}^D z_j^{-1/\theta})^\theta$, $\theta\in\Theta=(0,1]$. Here, the parameter $\theta$ controls the dependence strength, with $\theta\to0$ and $\theta=1$ corresponding to perfect dependence and independence, respectively. This model was chosen for two main reasons: first, it is the simplest max-stable distribution, often used as a benchmark, that interpolates between perfect dependence and independence; and second, the full likelihood \eqref{density} can be efficiently computed in this case using a recursive algorithm \citep{Shi:1995a}, thus allowing us to compare $\hat\theta_{\rm SEM}$ and $\hat\theta$ in high dimensions. Further results for the popular and more realistic, but computationally demanding, spatial max-stable model known as Brown--Resnick model \citep{Kabluchko.etal:2009}, are reported in the Supplementary Material. All simulations presented below and in the Supplementary Material were performed in parallel on the KAUST Cray XC40 supercomputer Shaheen II.

\begin{figure}[t!]
\centering
\includegraphics[width=\linewidth]{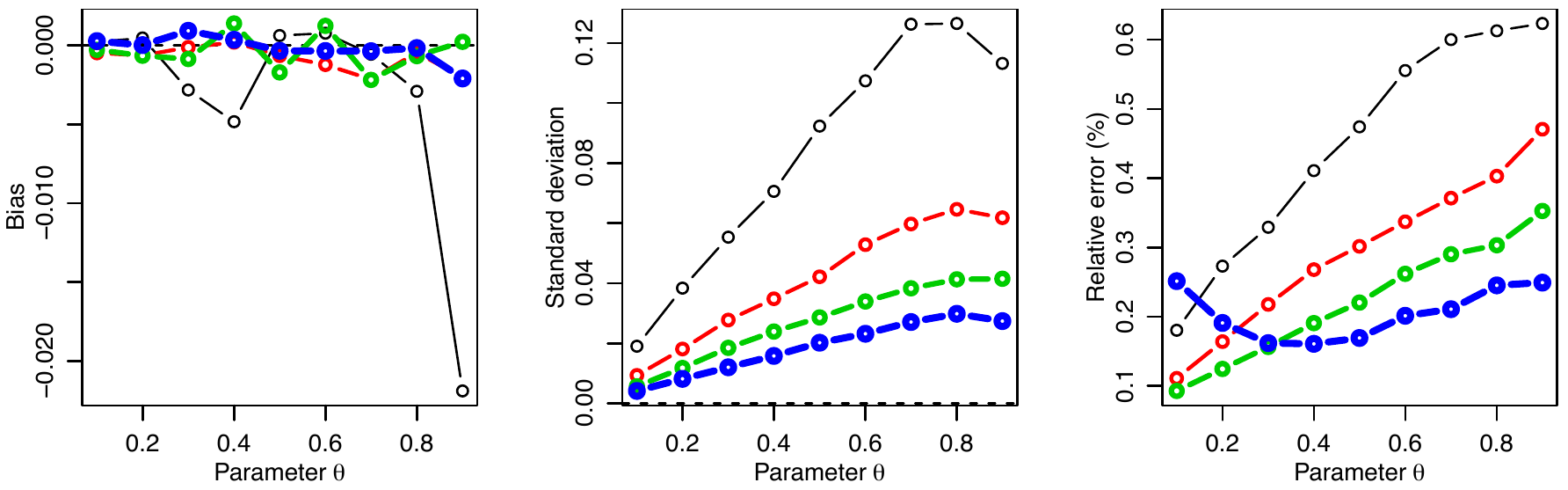}
\caption{Performance of the estimator $\hat\theta_{\rm SEM}$: bias (left), standard deviation (middle) and relative error in $\%$ (right), for the logistic model with $\theta=0.1,\ldots,0.9$ in dimension $D=2$ (thinnest black), $5$ (thin red), $10$ (thick green) and $20$ (thickest blue), based on $20$ independent temporal replicates. The number of iterations for the Expectation-Maximisation algorithm was set to $R=30$, averaging the last $5$ iterations, and the number of iterations of the underlying Gibbs sampler was set to $110\times D$ (thinned by a factor $D$, after a burn-in of $10\times D$). The initial value was set to $\theta_0=0.6$.}\label{Fig:results1}
\end{figure}

We first investigated the performance of the estimator $\hat\theta_{\rm SEM}$ under different scenarios. We considered dimensions $D=2,5,10,20$, with $20$ independent temporal replicates, and $\theta=0.1,\ldots,0.9$ (strong to weak dependence). Setting the initial value to $\theta_0=0.6$, we chose $R=30$ iterations for the Expectation-Maximisation algorithm, averaging the last $5$ iterations, and took $110\times D$ iterations for the underlying Gibbs sampler. Following simulations reported in the Supplementary Material, we discarded the first $10\times D$ iterations as burn-in, and thinned the Markov chain by a factor $D$, in order to keep $N=100$ roughly independent partitions $\pi_i$ to compute \eqref{StEstep}. We repeated the experiment $1024$ times to estimate the bias, ${\rm B}$, standard deviation, ${\rm SD}$, root mean squared error, ${\rm RMSE}=({\rm B}^2+{\rm SD}^2)^{1/2}$, and relative error with respect to the maximum likelihood estimator, ${\rm RE}=E|(\hat\theta_{\rm SEM}-\hat\theta)/\hat\theta|$. Figure~\ref{Fig:results1} reports the results. As expected, the bias is negligible compared to the standard deviation, and the latter decreases with increasing dimension $D$ but increases as the data approach independence ($\theta\to1$). The RMSE (not shown) is almost only determined by the standard deviation. The relative error is always very small (uniformly less than about $0.6\%$), and it decreases for the most part with $D$ and also with $N$ as suggested by further unreported simulations.

\begin{figure}[t!]
\centering
\includegraphics[width=\linewidth]{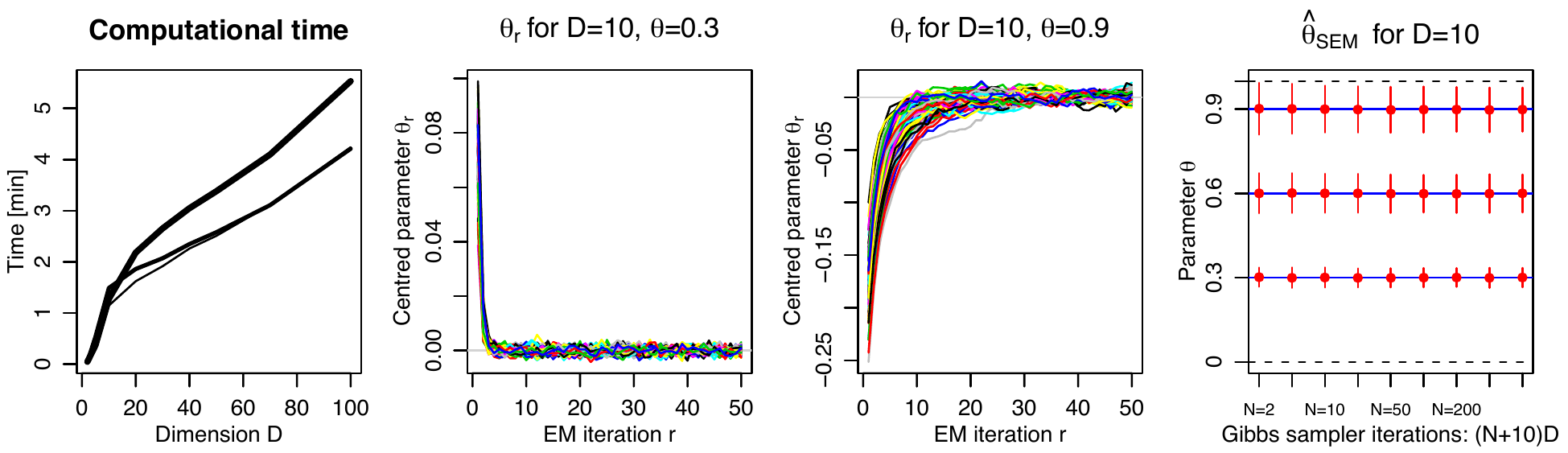}
\caption{Left: Computational time for computing $\hat\theta_{\rm SEM}$ for the logistic model, as function of dimension $D$, for $\theta=0.3$ (thin), $0.6$ (medium), $0.9$ (thick). We used $20$ temporal replicates, $30$ Expectation-Maximisation iterations, and $110\times D$ iterations for the underlying Gibbs sampler. Middle panels: parameter values $\theta_r$ as function of iteration $r=1,\ldots,50$, centred by the average over iterations $30$--$50$, for $100$ independent runs in dimension $D=10$. True values were set to $\theta=0.3$ (second panel) and $0.9$ (third panel), and the initial value was set to $\theta_0=0.6$. Right: Mean of estimated parameters $\hat\theta_{\rm SEM}$ (red dots) with $95\%$ confidence intervals for $\theta=0.3,0.6,0.9$, dimension $D=10$, $30$ Expectation-Maximisation iterations, and $(N+10)\times D$ Gibbs sampler iterations with $N=2,5,10,20,50,100,200,500,1000$ (x-axis). In all simulations, we always thinned the underlying Markov chains by a factor $D$ after discarding a burn-in of $10\times D$ iterations.}\label{Fig:results2}
\end{figure}

We now turn our attention to the computational efficiency of the stochastic Expectation-Maximisation algorithm. Considering dimensions up to $D=100$ under the exact same setting as before, the leftmost panel of Fig.~\ref{Fig:results2} shows that it takes on average $5$--$6$ minutes to compute $\hat\theta_{\rm SEM}$ (on a single core with 2.3GHz) when $D=100$ and $\theta=0.9$. Recall that, according to \citet{Castruccio.etal:2016}, a direct evaluation of the likelihood \eqref{density} is not possible in dimensions greater than $D=12$ or $13$, thus this result is a big improvement over the current existing methods. The computational time appears to be roughly linear with $D$, which is due to the number of iterations of the Gibbs sampler set proportional to $D$. However, for more complex models such as the Brown--Resnick model, the computational time is significantly larger; see the Supplementary Material. Therefore, it makes sense to seek strategies to reduce the computational burden. One possibility is to tune the number of iterations of the stochastic Expectation-Maximisation algorithm. To investigate its speed of convergence, the two middle panels of Fig.~\ref{Fig:results2} show the sample path $r\mapsto \theta_r$ for the logistic model as a function of the iteration $r=1,\ldots,50$, centred by the average over iterations $30$--$50$ for $100$ independent runs in dimension $D=10$. The true values were set to $\theta=0.3$ (second panel) and $0.9$ (third panel), and the initial value was set to $\theta_0=0.6$. The convergence is quite fast when $\theta=0.3$, requiring about $5$ iterations, but when $\theta=0.9$, it takes between $15$ and $25$ iterations. Simulations reported in the Supplementary Material suggest that although the Brown--Resnick has $p=2$ parameters, the number of iterations needed for the algorithm to converge is roughly the same. Another possibility to reduce the computational time is to play with the number of iterations of the underlying Gibbs sampler, which is the main computational bottleneck of this approach. The rightmost panel of Fig.~\ref{Fig:results2} displays estimated parameters in dimension $D=10$ with associated $95\%$ confidence intervals for $\theta=0.3,0.6,0.9$, using $30$ Expectation-Maximisation iterations and $(N+10)\times D$ iterations for the underlying Gibbs sampler with $N=2,5,10,20,50,100,200,500,1000$. Again, we discarded the first $10\times D$ iterations as burn-in and we thinned the resulting chain by a factor $D$. Surprisingly, the distribution of $\hat\theta_{\rm SEM}$ is almost stable for all $N\geq2$, suggesting that the number of Gibbs iterations does not need to be very large for accurate estimation. Overall, significant computational savings can be achieved without any loss of accuracy, by suitably choosing $R$ (Expectation-Maximisation iterations) and $N$ (Gibbs iterations).

\section{Discussion}\label{discussion}
To resolve the problem of inference for max-stable distributions and processes, we have proposed a stochastic Expectation-Maximisation algorithm, which does not fix the underlying partition, but instead, treats it as a missing observation, and integrates it out. The beauty of this approach is that it combines statistical and computational efficiency in high dimensions, and it does not suffer from misspecification entailed by lack of convergence of the partition. As a proof of concept, we have validated the methodology by simulation based on the logistic model, and we have shown that in this case it is easy to make inference beyond dimension $D=100$ in just a few minutes. In the Supplementary Material, we have also provided results for the popular Brown--Resnick spatial max-stable model. In this case, our full likelihood inference approach can handle dimensions up to about $D=20$ in a reasonable amount of time. The difficulty resides in the computation of high-dimensional multivariate Gaussian distributions needed for $V$ and $V_{\tau_i}$. Unbiased Monte Carlo estimates of these quantities can be obtained, and \citet{Thibaud.etal:2016} and \citet{deFondeville.Davison:2018} suggest using crude approximations to reduce the computational time while maintaining accuracy; see also \citet{Genton.etal:2018}. Our method is not limited to these two models and could potentially be applied to any max-stable model for which the exponent function $V$ and its partial derivatives $V_{\tau_i}$ are available. The main computational bottleneck of our approach is that we need to generate a Gibbs sampler for each independent temporal replicate of the process. Fortunately, as this setting is embarrassingly parallel, we may thus easily take advantage of available distributed computing resources. Finally, there is a large volume of literature on the stochastic Expectation-Maximisation algorithm, and it might be possible to devise automatic stopping criteria and adaptive schemes for the Gibbs sampler to further speed up the algorithm \citep{Booth.Hobert:1999}.

%\section*{Acknowledgement}
%This research was supported by...

\section*{Supplementary material}
Supplementary Material available online includes an alternative derivation of the Stephenson--Tawn likelihood based on a Poisson point process argument using the extremal functions, details on the calculation of the Poisson point process intensity for various max-stable models, details on the underlying Gibbs sampler, and further simulation results for the Brown--Resnick model.%, and details about the asymptotic behavior of the stochastic Expectation-Maximisation estimator.
%\section*{Acknowledgement}
%This research was supported by...

\newpage
\appendix
\section{Supplementary material}\label{SM}
%Supplementary Material available online includes an alternative derivation of the Stephenson--Tawn likelihood based on a Poisson point process argument using the extremal functions and details on the calculation of the Poisson point process intensity for various max-stable models.

\subsection{Likelihood derivation via Poisson point process intensity}
In their original paper, \citet{Stephenson.Tawn:2005} derived the likelihood functions $g_{{\rm Full}}$ and $g_{{\rm ST}}$ by differentiating the cumulative distribution function
$$\pr\{Z(s_1)\leq z_1,\ldots,Z(s_D)\leq z_D\}=\exp\{-V(z_1,\ldots,z_D)\}.$$
Here, we propose a different approach based on the analysis of the Poisson point process representation of the max-stable process
\begin{equation}\label{eq:deHaan}
Z(s)=\sup_{j\geq 1} W_j(s)/P_j,\quad s\in\calS.
\end{equation}
Introducing the functions $\varphi_j=W_j/P_j$, $j=1,2,\ldots$, the point process $\Phi=\{\varphi_j, j\geq 1\}$ is a Poisson point process on the space of nonnegative functions defined on $\mathcal{S}$. The max-stable process $Z$ appears as the pointwise maximum of the functions in $\Phi$. \citet{Dombry.EyiMinko:2013} showed that for all sites $s\in\mathcal{S}$, there almost surely exists a unique function in $\Phi$ that reaches the maximum $Z(s)$ at $s$. This function is called the extremal function at $s$ and denoted by $\varphi_s^+$. Clearly, $Z(s)=\varphi_s^+(s)$. 

Given $D$ sites $s_1,\ldots,s_D\in\calS$, there can be repetitions within the extremal functions $\varphi_{s_1}^+,\ldots,\varphi_{s_D}^+$, meaning that the maximum at different sites $s_{j_1},s_{j_2}$, can arise from the same extremal event. The notion of hitting scenario accounts for such possible repetitions. It is defined as the random partition  $\pi=\{\tau_1,\ldots,\tau_{|\pi|}\}$ (of size $|\pi|$) of $\{1,\ldots,D\}$ such that the two indices $j_1$ and $j_2$ are in the same block if and only if the extremal functions at $s_{j_1}$ and $s_{j_2}$ are equal. Here $k$ denotes the number of blocks of the partition $\pi$ and is equal to the number of different functions in $\Phi$ reaching the maximum $Z(s)$ for some point $s\in\{s_1,\ldots,s_D\}$. Within the block $\tau_i$, all the points $s_j$, $j\in\tau_i$, share the same extremal function that will hence be denoted by $\varphi_{\tau_i}^+$.

The joint distribution of the hitting scenario $\pi=\{\tau_1,\ldots,\tau_{|\pi|}\}$ and extremal functions $\{\varphi_{\tau_1}^+,\ldots,\varphi_{\tau_{|\pi|}}^+\}$ was derived by \citet{Dombry.EyiMinko:2013}. The max-stable observations $Z(s_1),\ldots,Z(s_D)$ relate to the hitting scenario and extremal functions via the simple equation $Z(s_j)=\varphi_{\tau_i}^+(s_j)$ for $j\in\tau_i$. In this way, we can deduce the joint distribution of the partition $\pi=\{\tau_1,\ldots,\tau_{|\pi|}\}$ and max-stable observations $Z(s_1),\ldots,Z(s_D)$, i.e., the Stephenson--Tawn likelihood $g_{{\rm ST}}$. Marginalising out the random partition, we deduce the full likelihood $g_{{\rm Full}}$.  

Suppose that the random vectors $\{W_j(s_1),\ldots,W_j(s_D)\}^{\T}$, $j\geq 1$, stemming from \eqref{eq:deHaan}, have a density $f_W$ with respect to the Lebesgue measure on $(0,+\infty)^D$. Then, the Poisson point process $\{\{\varphi_j(s_1),\ldots,\varphi_j(s_D)\}^{\T},j\geq 1\}$ on $(0,+\infty)^D$ has intensity
\begin{equation}\label{eq:intensity}
\lambda(z_1,\ldots,z_D)=\int_0^\infty f_W(z_1/r,\ldots,z_D/r)r^{-2-D}\mathrm{d}r.
\end{equation}
For clarity, we introduce some vectorial notation: let $\mathbf{s}=(s_1,\ldots,s_D)^{\T}$, $\mathbf{z}=(z_1,\ldots,z_D)^{\T}$, $Z(\mathbf{s})=\{Z(s_1),\ldots,Z(s_D)\}^{\T}$. For $\tau_i\subset\{1,\ldots,D\}$, $\tau_i^c$ denotes the complementary subset, and $\mathbf{z}_{\tau_i}$ and $\mathbf{z}_{\tau_i^c}$ are the subvectors of $\mathbf{z}$ obtained by keeping only the components from $\tau_i$ and $\tau_i^c$, respectively. Proposition 3 in \citet{Dombry.EyiMinko:2013} yields the following results:
\begin{itemize}
\item From the Poisson point process property, one can deduce the joint law of the hitting scenario and extremal functions:
$$\pr\{\pi=\{\tau_1,\ldots,\tau_{|\pi|}\},\varphi^+_{\tau_1}(\mathbf{s})=\mathrm{d}\mathbf{z}_1,\ldots,\varphi^+_{\tau_{|\pi|}}(\mathbf{s})=\mathrm{d}\mathbf{z}_k\}
=\exp\{-V(\max_{i=1}^{|\pi|}\mathbf{z}_i)\}\prod_{i=1}^{|\pi|} \lambda(\mathbf{z}_i)\mathrm{d}\mathbf{z}_i,$$
provided the partition associated to $\mathbf{z}_1,\ldots,\mathbf{z}_k$ is $\pi$; otherwise, this probability equals zero.
\item By definition of the extremal functions, one gets the joint law of the hitting scenario and max-stable observations:
\begin{equation}\label{eq:ST}
\pr\{\pi=\{\tau_1,\ldots,\tau_{|\pi|}\},Z(\mathbf{s})=\mathrm{d}\mathbf{z}\}=\exp\{-V(\mathbf{z})\}\left(\prod_{i=1}^{|\pi|} \int_{\mathbf{u}_{i}<\mathbf{z}_{\tau_i^c}}\lambda(\mathbf{z}_{\tau_i},\mathbf{u}_i)\mathrm{d}\mathbf{u}_i \right)\mathrm{d}\mathbf{z}.
\end{equation}
\item By integrating out the hitting scenario, one obtains the law of the max-stable observations:
\begin{equation}\label{eq:full}
\pr\{Z(\mathbf{s})=\mathrm{d}\mathbf{z}\}=\exp\{-V(\mathbf{z})\}\sum_{\pi\in\mathcal{P}_D}\left(\prod_{i=1}^{|\pi|} \int_{\mathbf{u}_{i}<\mathbf{z}_{\tau_i^c}}\lambda(\mathbf{z}_{\tau_i},\mathbf{u}_i)\mathrm{d}\mathbf{u}_i \right)\mathrm{d}\mathbf{z}.
\end{equation}
\end{itemize}
Equation \eqref{eq:ST} provides an alternative formula for the Stephenson--Tawn likelihood, $g_{{\rm ST}}$, based on the Poisson point process intensity, $\lambda$, while Equation \eqref{eq:full} is the max-stable full likelihood, $g_{{\rm Full}}$. Identifying the expressions \eqref{eq:ST} and \eqref{eq:full} above with (5) and (4) in the main paper, respectively, we can see that
\begin{equation}\label{partialder}
-\partial_{\tau_i}V(z_1,\ldots,z_D)=\int_{\mathbf{u}_{i}<\mathbf{z}_{\tau_i^c}}\lambda(\mathbf{z}_{\tau_i},\mathbf{u}_i)\mathrm{d}\mathbf{u}_i.
\end{equation}
This relates a partial derivative of the exponent function $V$ with a partial integral of the point process intensity $\lambda$. In particular, \eqref{partialder} implies that the intensity is the mixed derivative of the exponent function with respect to all arguments, i.e., 
\begin{equation}\label{intens}
\lambda(z_1,\ldots,z_D)=-{\partial^D\over \prod_{i=1}^D\partial z_i}V(z_1,\ldots,z_D).
\end{equation}
Furthermore, the function $V$ corresponds to the integrated intensity of the set $A=[0,\mathbf{z}]^c$, i.e.,  
\begin{equation*}\label{Vfun}
V(z_1,\ldots,z_D)=\Lambda([0,\mathbf{z}]^c)=\int_A\lambda(\mathbf{u})\mathrm{d}\mathbf{u}.
\end{equation*}

\subsection{Computing the Poisson point process intensity}
The intensity measure $\lambda$ is an important feature of max-stable models and can be computed for most popular models; see \citet{Dombry.etal:2013} for a derivation of $\lambda$ for the Brown--Resnick model \citep{Kabluchko.etal:2009} and \citet{Ribatet:2013b} for an expression of $\lambda$ for the extremal-$t$ model \citep{Opitz:2013a}. Partial integrals of $\lambda$ for these models may be found in \citet{Wadsworth.Tawn:2014} and \citet{Thibaud.Opitz:2015}, respectively. Using the relations \eqref{partialder} and \eqref{intens}, the intensity $\lambda$ and its partial integrals can be deduced for the \citet{Reich.Shaby:2012a} model from the expressions in the appendix of \citet{Castruccio.etal:2016}. 

Here, as a simple pedagogical illustration for many other multivariate or spatial max-stable models, we consider the multivariate logistic model, which we used in our simulation study. In this case, the function $V$ and its partial and full derivatives can readily be obtained by direct differentiation.

Recall that the exponent function for the logistic model is
$$V(z_1,\ldots,z_D\mid\theta)=\left(z_1^{-1/\theta}+\cdots+z_D^{-1/\theta}\right)^\theta,\quad \theta\in (0,1].$$
It is known that the multivariate counterpart of the spectral representation \eqref{eq:deHaan} for the logistic model is obtained by taking $W=(W_1,\ldots,W_D)^{\T}$ with independent and identically distributed Fr\'echet$(\beta,c_\beta)$ components, where $\beta=1/\theta$ and $c_\beta=1/\Gamma(1-1/\beta)$ are shape and scale parameters, respectively; see, for example, Proposition 6 in \citet{Dombry.etal:2016}. Then, 
$$
f_W(z_1,\ldots,z_D)=\prod_{i=1}^D \frac{\beta}{c_\beta}\left(\frac{z_i}{c_\beta}\right)^{-1-\beta}e^{-(z_i/c_\beta)^{-\beta}},
$$
and we deduce from Equation \eqref{eq:intensity} that
\begin{eqnarray*}
\lambda(z_1,\ldots,z_D)
&=&\int_{0}^\infty   \left[ \prod_{i=1}^D \frac{\beta}{c_\beta}\left(\frac{z_i}{rc_\beta}\right)^{-1-\beta}e^{-\{z_i/(rc_\beta)\}^{-\beta}} \right] r^{-2-D}\mathrm{d}r\\
&=&\frac{\Gamma(D-1/\beta)}{\beta }\left\{ \sum_{i=1}^D (z_i/c_\beta)^{-\beta}\right\}^{1/\beta-D}\prod_{i=1}^D \frac{\beta}{c_\beta}\left(\frac{z_i}{c_\beta}\right)^{-1-\beta}.  
\end{eqnarray*}
Similar computations entail
\begin{eqnarray*}
\int_{\mathbf{u}_{i}<\mathbf{z}_{\tau_i^c}}\lambda(\mathbf{z}_{\tau_i},\mathbf{u}_i)\mathrm{d}\mathbf{u}_i&=&\int_0^\infty \left[\prod_{j\in\tau_i} \frac{\beta}{c_\beta}\left(\frac{z_j}{rc_\beta}\right)^{-1-\beta}e^{-\{z_j/(rc_\beta)\}^{-\beta}} \right]\times\left[ \prod_{j \in \tau_i^c} e^{-\{z_j/(rc_\beta)\}^{-\beta}} \right]r^{-2-|\tau_i|}\mathrm{d}r\\
&=&\left\{\prod_{j\in\tau_i} \frac{\beta}{c_\beta}\left(\frac{z_j}{c_\beta}\right)^{-1-\beta}\right\}\int_0^\infty e^{-\sum_{j=1}^D\{z_j/(rc_\beta)\}^{-\beta}}  r^{\beta|\tau_i|-2}\mathrm{d}r\\
&=& \frac{\Gamma(|\tau_i|-1/\beta)}{\beta} \left\{\sum_{j=1}^D (z_j/c_\beta)^{-\beta}\right\}^{1/\beta-|\tau_i|}\prod_{j\in\tau_i} \frac{\beta}{c_\beta}\left(\frac{z_j}{c_\beta}\right)^{-1-\beta}\\
&=& \beta^{|\tau_i|-1}\frac{\Gamma(|\tau_i|-1/\beta)}{\Gamma(1-1/\beta)} \,\left(\sum_{j=1}^D z_j^{-\beta}\right)^{1/\beta-|\tau_i|} \prod_{i\in\tau_i} z_j^{-1-\beta},\\
\end{eqnarray*}
where, for the first equality, we used
$$\prod_{j\in\tau_i^c} \int_0^{z_j}\frac{\beta}{c_\beta}\left(\frac{u_j}{rc_\beta}\right)^{-1-\beta}e^{-\{u_j/(rc_\beta)\}^{-\beta}}\mathrm{d}u_j =\prod_{i\in\tau_i^c} r e^{-\{z_j/(rc_\beta)\}^{-\beta}}. $$

\subsection{Details on the underlying Gibbs sampler}\label{sec:Gibbs}
The Gibbs sampler proposed by \citet{Dombry.etal:2013} is designed to draw an ergodic sequence of partitions $\pi_1,\ldots,\pi_N$ conditional on the observed max-stable data $z=(z_1,\ldots,z_D)^{\T}$, i.e., from the discrete distribution $g(\pi\mid z,\theta)$, where $\theta\in\Theta\subset\Real^p$ is the parameter vector characterizing the max-stable dependence structure. One has
\begin{eqnarray}
g(\pi\mid z,\theta)&=&{g_{\rm ST}(\pi,z\mid\theta)\over g_{\rm Full}(z\mid \theta)}={\exp\{-V(z\mid\theta)\}\prod_{i=1}^{|\pi|}\{-V_{\tau_i}(z\mid\theta)\}\over \exp\{-V(z\mid\theta)\}\sum_{\pi\in\calP_D}\prod_{i=1}^{|\pi|}\{-V_{\tau_i}(z\mid \theta)\}}\nonumber\\
&=& {\prod_{i=1}^{|\pi|}\{-V_{\tau_i}(z\mid\theta)\}\over \sum_{\pi\in\calP_D}\prod_{i=1}^{|\pi|}\{-V_{\tau_i}(z\mid \theta)\}}\propto \prod_{i=1}^{|\pi|}\{-V_{\tau_i}(z\mid\theta)\}.\label{eq:Gibbs}
\end{eqnarray}
The normalizing constant in the denominator of \eqref{eq:Gibbs} is computationally demanding to compute as it involves the sum over all partitions. Nevertheless, the Gibbs sampler of \citet{Dombry.etal:2013} provides a way to construct a Markov chain whose stationary distribution is $g(\pi\mid z,\theta)$, while avoiding the computation of the normalizing constant. Let $\pi_t=\{\tau_{t;1},\ldots,\tau_{t;|\pi_t|}\}\in\calP_D$ be the partition at the $t$th iteration of the Gibbs sampler. The idea of the Gibbs sampler is to sample the next partition $\pi_{t+1}=\{\tau_{t+1;1},\ldots,\tau_{t+1;|\pi_{t+1}|}\}\in\calP_D$ by keeping all but one component fixed. Let $\ell\in\{1,\ldots,D\}$ be the component to be updated, and let $\pi_t^{-\ell}$ and $\pi_{t+1}^{-\ell}$ denote the partitions $\pi_t$ and $\pi_{t+1}$, respectively, with the $\ell$th component removed. We update the partition $\pi_t$ by modifying the (randomly chosen) $\ell$th component using the full conditional distribution
\begin{equation}\label{eq:Gibbs2}
g(\pi_{t+1}\mid\pi_{t+1}^{-\ell}=\pi_t^{-\ell},z,\theta)\propto {\prod_{i=1}^{|\pi_{t+1}|}\{-V_{\tau_{t+1;i}}(z\mid\theta)\}\over \prod_{i=1}^{|\pi_t|}\{-V_{\tau_{t;i}}(z\mid\theta)\}}.
\end{equation}
The combinatorial explosion is avoided, because the number of possible updates $\pi_{t+1}$ such that $\pi_{t+1}^{-\ell}=\pi_t^{-\ell}$ is at most $|\pi_t|+1$. Moreover, as we update only one component at a time, many terms in the ratio \eqref{eq:Gibbs2} cancel out, and at most four of them need to be computed, which makes it computationally attractive. However, for the same reason, the resulting partitions will also be heavily dependent, and so, intuitively, we should take the number of Gibbs iterations to be roughly proportional to the dimension $D$ and thin the Markov chain by a factor $D$ to get approximately independent (or weakly dependent) partitions. A suitable burn-in should also be specified to ensure that the Markov chain has appropriately converged to its stationary distribution.

\begin{figure}[t!]
\centering
\includegraphics[width=\linewidth]{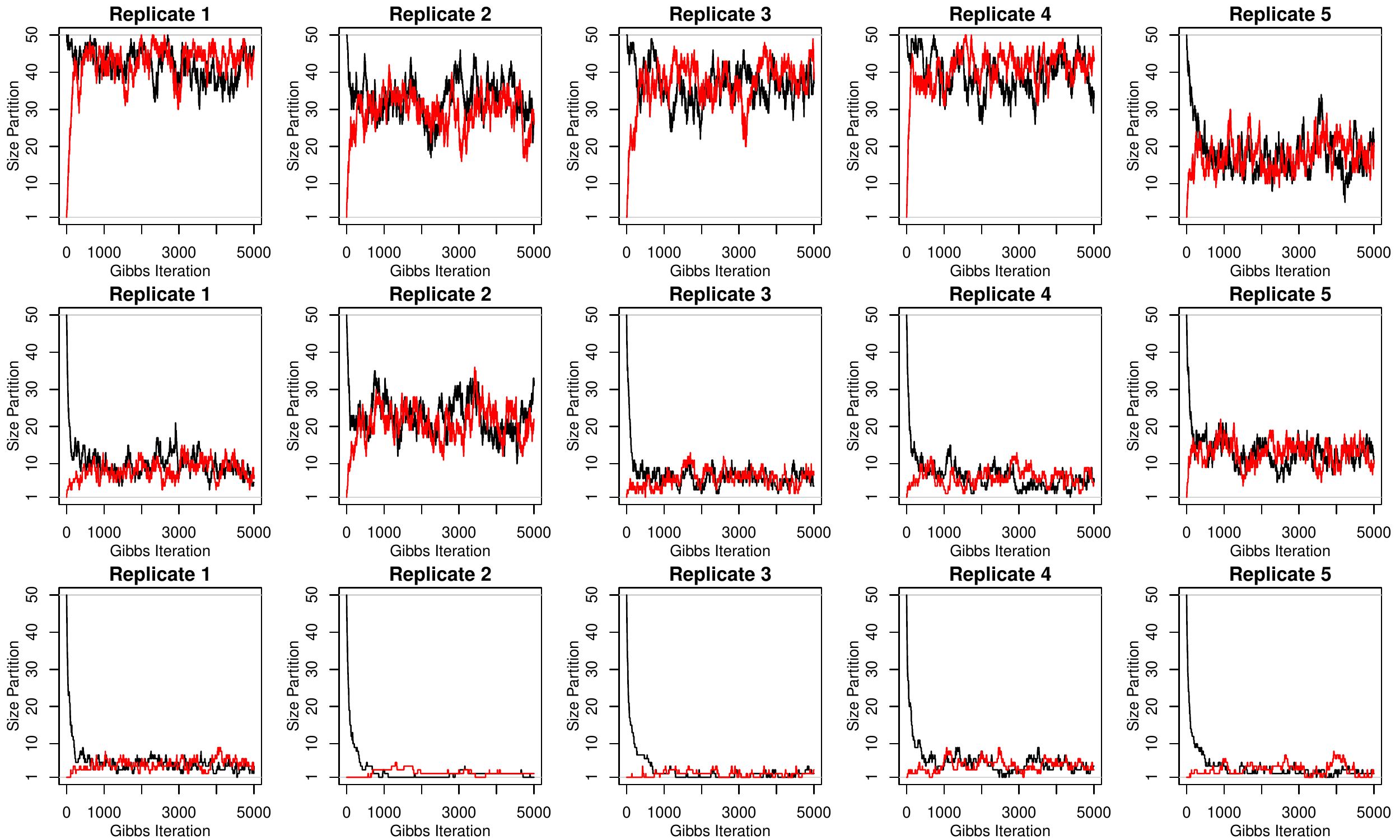}
\caption{Trace plots of the sizes of partitions obtained from the Gibbs samplers for each of the five replicates (columns). We considered the logistic model in dimension $D=50$ with parameter $\theta=0.9,0.6,0.3$ (top to bottom rows). Initial partitions were taken as $\{\{1\},\ldots,\{D\}\}$ (black) and $\{\{1,\ldots,D\}\}$ (red). $5000$ iterations were performed.}\label{fig:GibbsLog}
\end{figure}

In order to assess the number of iterations required for the Gibbs sampler to converge (i.e., the burn-in), we considered the logistic model defined by its exponent function $V(z_1,\ldots,z_D\mid\theta)=(z_1^{-1/\theta}+\cdots+z_D^{-1/\theta})^\theta$, $\theta\in (0,1]$. We generated five independent copies of a logistic random vector in dimension $D=50$, and considered the cases $\theta=0.9,0.6,0.3$ (weak to strong dependence). For each dataset, we ran five Gibbs samplers (one per independent replicate) for $5000$ iterations. To easily visualize the resulting Markov chains and assess convergence, we display in Fig.~\ref{fig:GibbsLog} trace plots of the sizes of partitions along the different Markov chains. The initial partitions were taken as $\{\{1\},\ldots,\{D\}\}$ (of size $D=50$), which reflects weak dependence scenarios, and $\{\{1,\ldots,D\}\}$ (of size one), which reflects strong dependence scenarios.
In all cases, we can see that the Gibbs sampler converges rather quickly, and that it is enough to discard a burn-in of about $10\times D=500$ iterations.

\begin{figure}[t!]
\centering
\includegraphics[width=\linewidth]{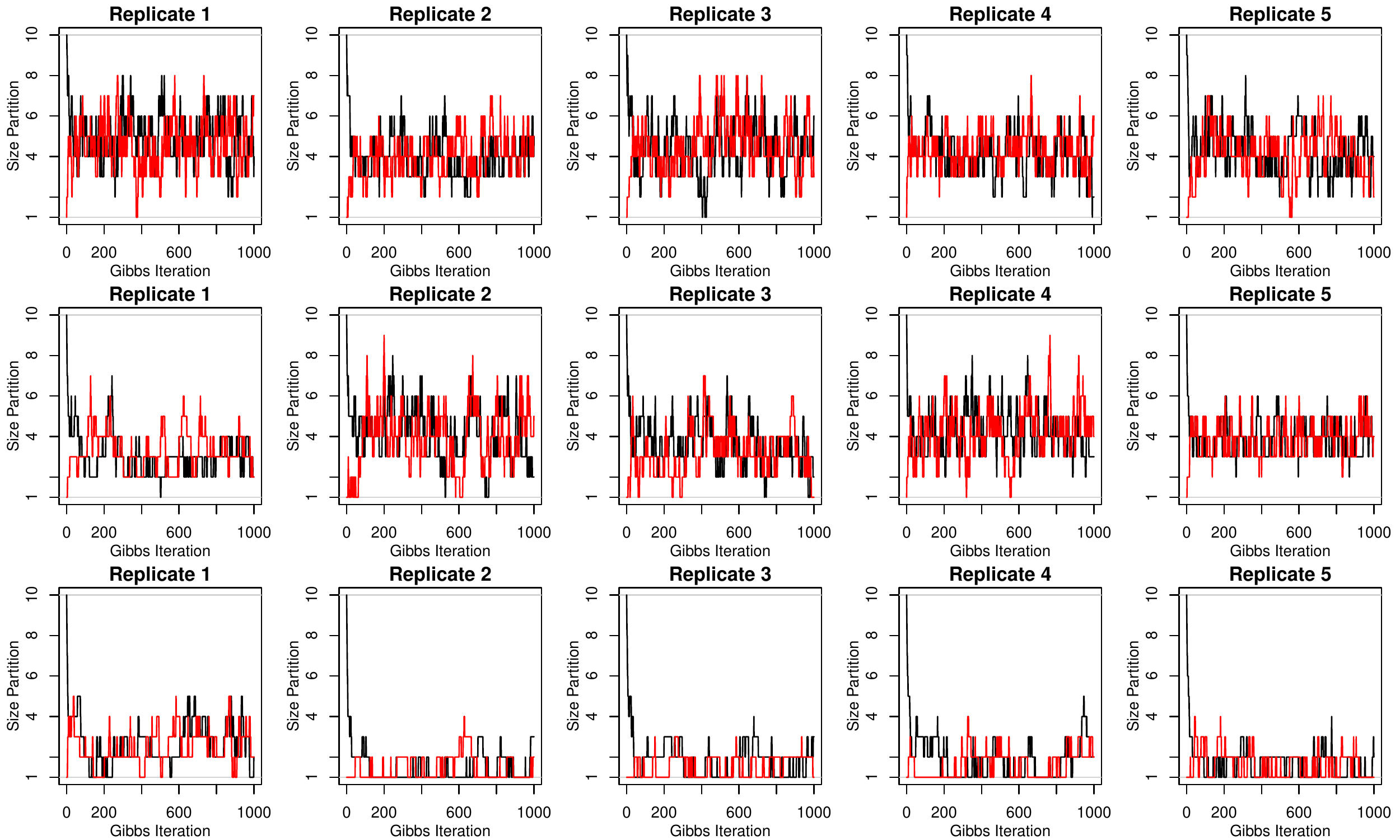}
\caption{Trace plots of the sizes of partitions obtained from the Gibbs samplers for each of the five replicates (columns). We considered the Brown--Resnick model at $D=10$ sites in $[0,1]^2$ with semi-variogram $\gamma(h)=(h/\lambda)^\nu$ and parameters $\lambda=0.5,1,1.5$ (top to bottom rows) with $\nu=1.5$. Initial partitions were taken as $\{\{1\},\ldots,\{D\}\}$ (black) and $\{\{1,\ldots,D\}\}$ (red). $1000$ iterations were performed.}\label{fig:GibbsBR}
\end{figure}

To validate such results for another max-stable model, we did the same experiment for the Brown--Resnick model \citep{Kabluchko.etal:2009}, defined by taking $W_j(s)=\exp\{\varepsilon_j(s)-\gamma(s)\}$ in \eqref{eq:deHaan}, where the terms $\varepsilon_j(s)$ are independent copies of $\varepsilon(s)$, which is an intrinsically stationary Gaussian process with zero mean and variogram $2\gamma(h)={\rm var}\{\varepsilon(s)-\varepsilon(s+h)\}$ such that $\varepsilon(0)=0$ almost surely. Using the exact simulation algorithm of \citet{Dombry.etal:2016}, we simulated five independent replicates of the Brown--Resnick with semi-variogram $\gamma(h)=(h/\lambda)^\nu$, where $\lambda>0$ and $\nu\in(0,2]$ are the range and smoothness parameters, respectively, at $D=10$ randomly generated sites $s_1,\ldots,s_{10}\in[0,1]^2$. We considered the cases $\lambda=0.5,1,1.5$ (short to long range dependence) with $\nu=1.5$. For each dataset, we ran five Gibbs samplers (one per independent replicate) for $1000$ iterations. Figure~\ref{fig:GibbsBR} shows the trace plots of the sizes of partitions along the different Markov chains. As before, the initial partitions were taken as $\{\{1\},\ldots,\{D\}\}$ (of size $D=10$) and $\{\{1,\ldots,D\}\}$ (of size one).
As concluded for the logistic model, we can see that the Gibbs sampler converges quickly and that about $10\times D=100$ iterations are enough for the algorithm to converge in all cases. 

These results suggest to discard the first $10\times D$ iterations as burn-in, and to thin the resulting Markov chains by a factor $D$ to obtain approximately (conditionally) independent partitions. With this setting, the initial partition has negligible impact on the results. Alternatively, another natural option could be to initialize the partition randomly from its unconditional distribution, which can be easily obtained from an unconditional simulation of the max-stable distribution. This could potentially provide further computational savings by reducing the time it takes for the Gibbs sampler to converge (thus reducing the burn-in).

\subsection{Simulation results for the Brown--Resnick model}
\begin{figure}[t!]
\centering
\includegraphics[width=\linewidth]{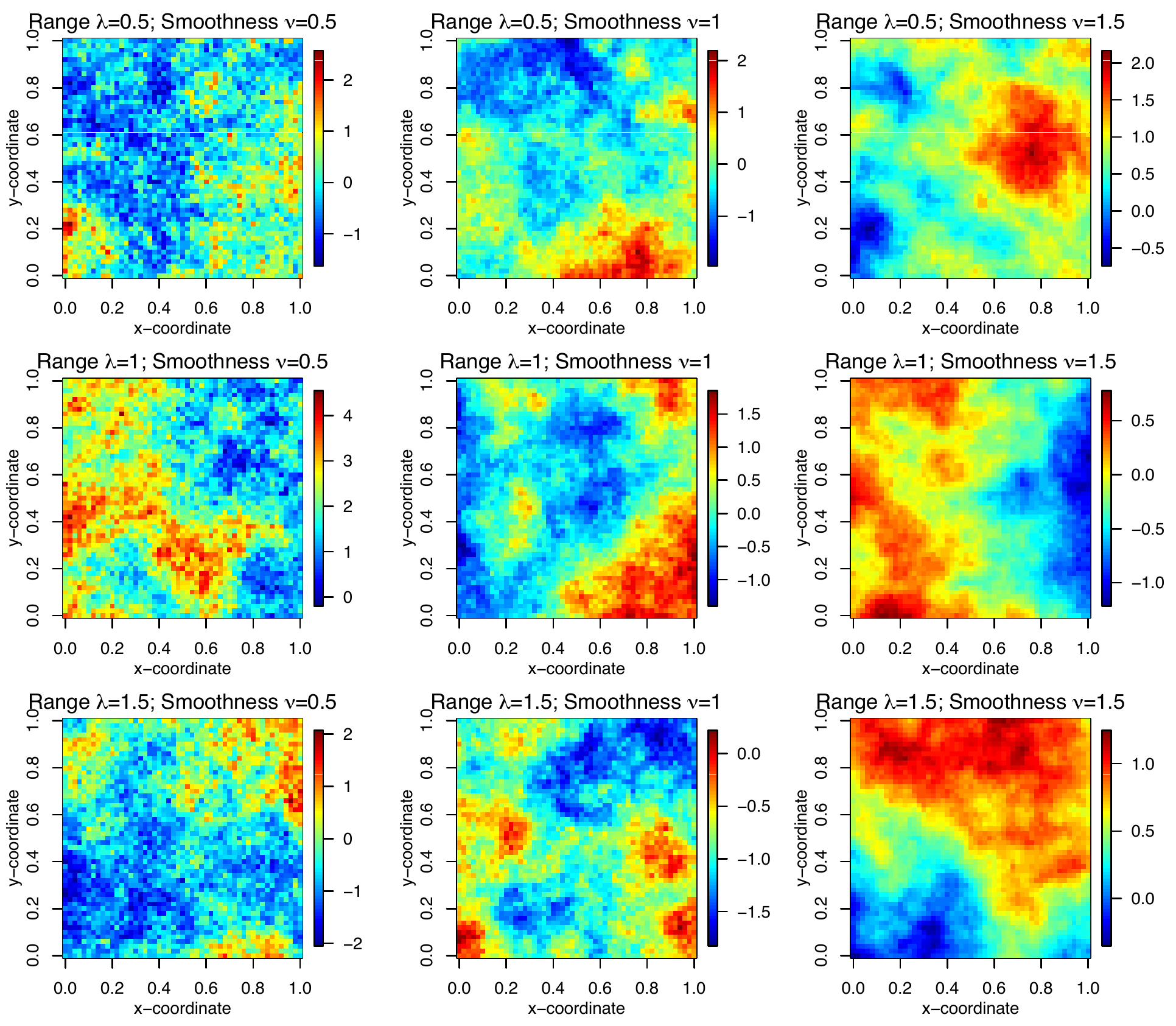}
\caption{Realizations from the Brown--Resnick model on $[0,1]^2$, with semi-variogram $\gamma(h)=(h/\lambda)^\nu$, and parameter values taken according to Table~\ref{tab:scenariosBR}, covering short to long range dependent processes (top to bottom) and rough to smooth processes (left to right). Realizations are displayed on standard Gumbel margins.}\label{fig:realizationsBR}
\end{figure}

We now provide further simulation results for the Brown--Resnick model. We follow the definition given in \S\ref{sec:Gibbs} using the semi-variogram $\gamma(h)=(h/\lambda)^\nu$, where $\lambda>0$ and $\nu\in(0,2]$ are the range and smoothness parameters, and we consider the scenarios displayed in Table~\ref{tab:scenariosBR}. Realizations for each scenario are illustrated in Fig.~\ref{fig:realizationsBR}.  
\begin{table}[t!]
\centering
\caption{Scenarios considered for the simulation study based on the Brown--Resnick model with semi-variogram $\gamma(h)=(h/\lambda)^\nu$.}\label{tab:scenariosBR}
\begin{tabular}{ccccccccc}
\multicolumn{9}{c}{Scenarios (i.e., parameter configurations $\lambda;\nu$)}\\
1 & 2 & 3 & 4 & 5 & 6 & 7 & 8 & 9 \\ \hline
$0.5;0.5$ & $0.5;1.0$ & $0.5;1.5$ & $1.0;0.5$ & $1.0;1.0$ & $1.0;1.5$ & $1.5;0.5$ & $1.5;1.0$ & $1.5;1.5$
\end{tabular}
\end{table}

In order to assess the performance of the stochastic Expectation-Maximisation estimator in each scenario of Table~\ref{tab:scenariosBR}, we simulated in each case $10$ independent copies of the Brown--Resnick model at $D=10$ randomly generated sites in $[0,1]^2$, and then estimated the range and smoothness parameters. We used (i) the stochastic Expectation-Maximisation estimator $\hat\theta_{\rm SEM}=(\hat\lambda_{\rm SEM},\hat\nu_{\rm SEM})^{\T}$ based on $60\times D=600$ Gibbs iterations in total, then thinning by a factor $D=10$, after discarding a burn-in of $10\times D=100$ iterations to follow the results of \S\ref{sec:Gibbs}; and (ii) a pairwise likelihood estimator $\hat\theta_{\rm PAIR}=(\hat\lambda_{\rm PAIR},\hat\nu_{\rm PAIR})^{\T}$ \citep[see, e.g.,][]{Padoan.etal:2010,Huser.Davison:2013a}, which maximizes the pairwise likelihood constructed by combining the likelihood contributions from all ${10 \choose 2}=45$ pairs of sites together with equal weight. We repeated this experiment $1024$ times to compute performance metrics, such as the root mean squared error of parameter estimates.

\begin{figure}[t!]
\centering
\includegraphics[width=\linewidth]{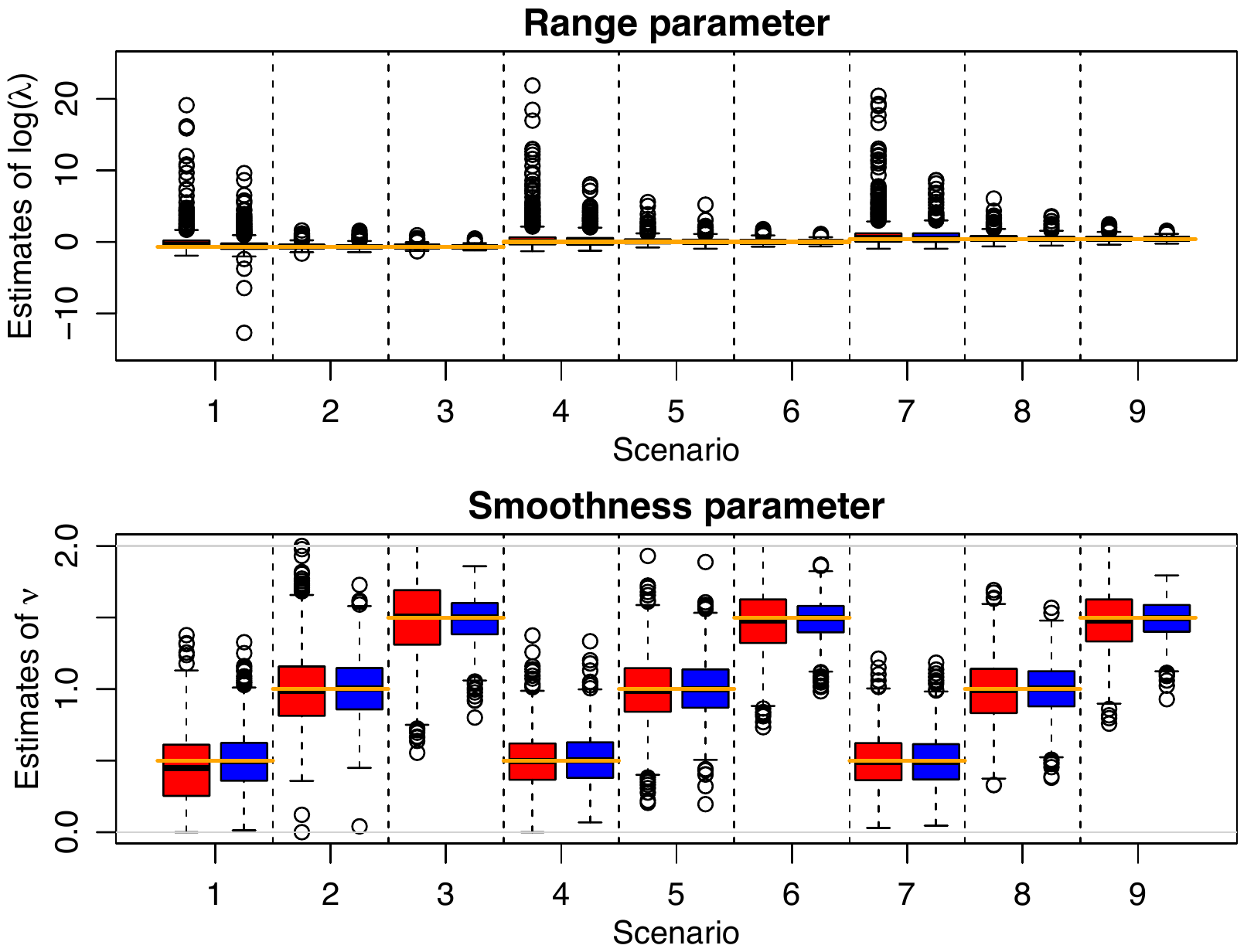}
\caption{Boxplots of estimates of $\log(\lambda)$ (top) and $\nu$ (bottom) for each scenario in Table~\ref{tab:scenariosBR} based on the Brown--Resnick model with semi-variogram $\gamma(h)=(h/\lambda)^\nu$, simulated at $D=10$ random sites in $[0,1]^2$, with 10 independent replicates. Left (red) and right (blue) boxplots correspond to $\hat\theta_{\rm PAIR}=(\hat\lambda_{\rm PAIR},\hat\nu_{\rm PAIR})^{\T}$ and $\hat\theta_{\rm SEM}=(\hat\lambda_{\rm SEM},\hat\nu_{\rm SEM})^{\T}$, respectively. Each boxplot summarizes the variability of parameter estimates based on $1024$ simulations. Five estimates reaching up to $\log(\hat\lambda_{\rm PAIR})\approx40$ were omitted in Scenario 1 for visibility purposes. Orange horizontal segments are the true values.}\label{fig:boxplotsBR}
\end{figure}
Figure~\ref{fig:boxplotsBR} displays boxplots of estimated parameters for each scenario. Both stochastic Expectation-Maximisation and pairwise likelihood estimators seem to work well overall with a very low bias, although the variability is in some cases very high, due to the tricky estimation exercise with only $10$ replicates in dimension $D=10$. Nevertheless, the inter-quartile range appears to be quite moderate in all cases. The stochastic Expectation-Maximisation estimator appears clearly superior to the pairwise likelihood estimator in the cases we have considered, as it fully utilizes the information available in the data. To investigate this more in depth, Table~\ref{tab:ARE_BR} reports relative efficiencies of the pairwise likelihood estimator with respect to the stochastic Expectation-Maximisation estimator, defined as the ratio between the corresponding root mean squared errors (calculated from the 1024 replicates). 
\begin{table}[t!]
\centering
\caption{Relative efficiencies ($\%$) for the estimates of $\log(\lambda)$ (left) and $\nu$ (right) based on the pairwise likelihood estimator with respect to the stochastic Expectation-Maximisation estimator. Simulations were based on the Brown--Resnick model with semi-variogram $\gamma(h)=(h/\lambda)^\nu$, simulated at $D=10$ random sites in $[0,1]^2$, with 10 independent replicates.}\label{tab:ARE_BR}
\begin{tabular}{r|ccc}
$\lambda/\nu$ & $\nu=0.5$ & $\nu=1.0$ & $\nu=1.5$ \\ \hline
$\lambda=0.5$ & $39;75$ & $94;80$ & $77;59$ \\
$\lambda=1.0$ & $57;92$ & $82;87$ & $73;62$ \\
$\lambda=1.5$ & $54;92$ & $79;82$ & $71;61$ 
\end{tabular}
\end{table}
The results suggest that the stochastic Expectation-Maximisation estimator has a much better performance overall, as expected. Moreover, such results are expected to improve in higher-dimensional settings, where the loss in efficiency of pairwise likelihood estimators is more significant.

\begin{figure}[t!]
\centering
\includegraphics[width=0.8\linewidth]{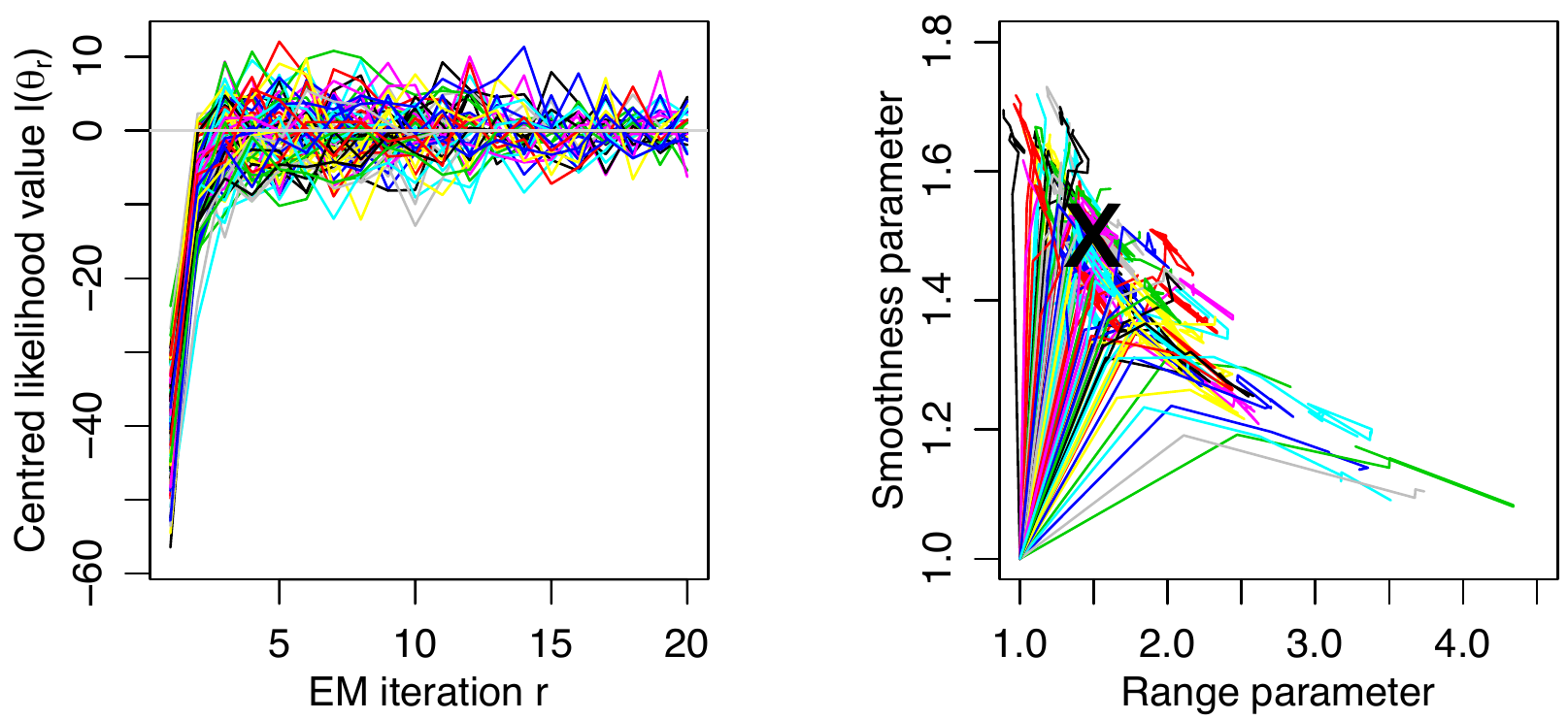}
\caption{\emph{Left:} likelihood value $l(\theta_r)$ plotted as a function of the Expectation-Maximisation iteration $r=0,1,\ldots,20$, for $100$ simulations based on the Brown--Resnick model with semi-variogram $\gamma(h)=(h/\lambda)^\nu$, simulated at $D=10$ random sites in $[0,1]^2$, with 10 independent replicates. The likelihood values were centred by their average over iterations 16--20 for visibility purposes. \emph{Right:} Trace of corresponding parameter values $\theta_r=(\lambda_r,\nu_r)^{\T}$, plotted for each Expectation-Maximisation iteration $r=0,1,\ldots,20$. The true values (black cross) are $\lambda=\nu=1.5$, while the initial values were taken to be $\lambda_0=\nu_0=1$.}\label{fig:EM_BR}
\end{figure}

We now desire to check the speed of convergence of the Expectation-Maximisation algorithm, similarly to the simulations that we did for the logistic model in the main paper. Figure~\ref{fig:EM_BR} displays the value of the likelihood (left panel) and the parameters (right panel) for each iteration $r=0,1,\ldots,20$ of the Expectation-Maximisation algorithm, for $100$ simulations performed in the exact same setting as above with true values chosen to be $\lambda=\nu=1.5$. The likelihood values were centred by their average over iterations 16--20 for visibility purposes. The plots show that the Expectation-Maximisation algorithm converges after roughly 5 iterations in this case, which is similar to the results obtained for the logistic model, recall the results in the main paper, despite the fact that the Brown--Resnick model has one more parameter ($p=2$). Hence, in practice, a small number of iterations could be chosen to speed up the algorithm.

The right panel of Fig.~\ref{fig:EM_BR} also reveals that the estimated range parameter is negatively correlated with the estimated smoothness parameter. This was expected as these two parameters have an opposing effect on the dependence strength, and it suggests that alternative orthogonal parametrizations might be preferable. We leave this problem for future research.

\begin{figure}[t!]
\centering
\includegraphics[width=0.6\linewidth]{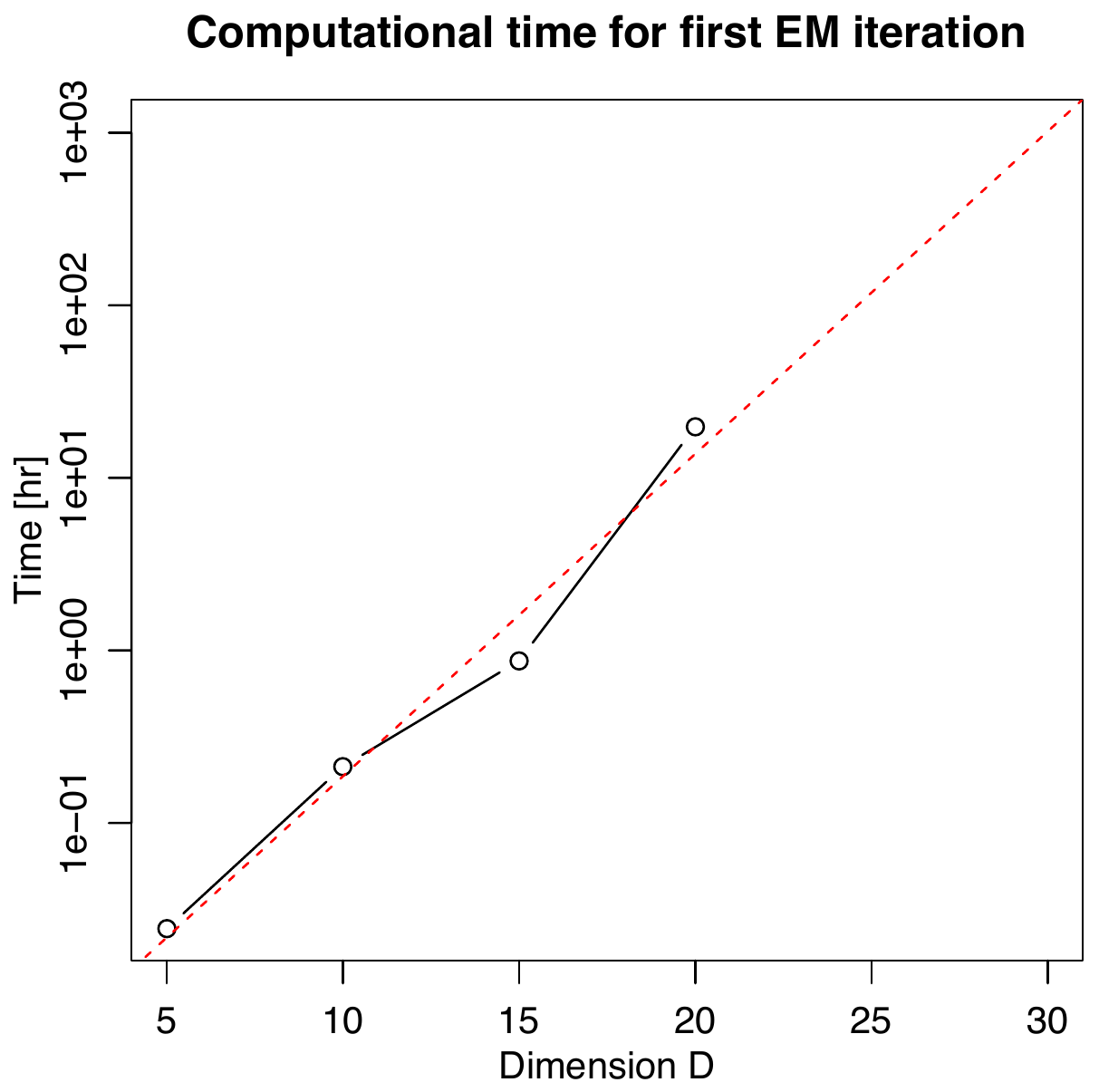}
\caption{Computational time (hr) plotted on a logarithmic scale (black dots), for the first Expectation-Maximisation iteration. The values were obtained by averaging $1024$ computing times based on the Brown--Resnick model with semi-variogram $\gamma(h)=(h/\lambda)^\nu$, $\lambda=\nu=1.5$, simulated at $D=5,10,15,20$ random sites in $[0,1]^2$, with 10 independent replicates. The dashed red line represents a projection for higher dimensions, based on a linear fit (on logarithmic scale).}\label{fig:time_BR}
\end{figure}

Finally, we investigate the scalability of the stochastic Expectation-Maximisation estimator for the Brown--Resnick model when the dimension $D$ increases. To assess this, we considered the same setting as above with $10$ independent replicates, same number of Gibbs iterations, true values set to $\lambda=\nu=1.5$, and we measured the computational time needed for the first Expectation-Maximisation iteration, in dimensions $D=5,10,15,20$. Figure~\ref{fig:time_BR} plots the computational time on a logarithmic scale, along with a projection for dimensions up to $D=30$. 
Unlike the logistic model, which has an explicit exponent function $V$ and partial derivatives $V_{\tau_i}$, the expressions for the Brown--Resnick model involve the multivariate Gaussian distribution in dimension up to $D-1$, whose computation is very demanding for large $D$. This significantly slows down the algorithm, whose complexity now appears to be increasing exponentially with $D$ (by contrast with the linear increase observed for the logistic model). More precisely, a single Expectation-Maximisation iteration takes on average approximately $1.5$min, $12.7$min, $52.2$min, and $19.8$hr for dimensions $D=5,10,15,20$, respectively. This precludes the use of the stochastic Expectation-Maximisation algorithm beyond dimension $D=20$, unless distributed computing resources are used to run all Gibbs sampler in parallel. Nevertheless, these results already represent a big advancement  towards full likelihood inference for max-stable processes, as \citet{Castruccio.etal:2016} argued that a direct calculation of the likelihood was impossible beyond dimension $D=12$ or $13$. As shown by \citet{deFondeville.Davison:2018}, speed-ups can be obtained by appropriately exploiting quasi-Monte Carlo techniques for the calculation of multivariate Gaussian distributions. Alternatively, we could also use hierarchical matrix decompositions \citep{Genton.etal:2018}.

A similar computational burden is expected for the extremal-$t$ model \citep{Opitz:2013a}, which relies on the computation of multivariate Student $t$ distributions, but a better computational efficiency should prevail for the \citet{Reich.Shaby:2012a} max-stable model, for which the expressions of the exponent function $V$ and its partial derivatives $V_{\tau_i}$ are available in explicit form; see the appendix of \citet{Castruccio.etal:2016}.

\baselineskip=24pt

\bibliographystyle{CUP}
\bibliography{paper-ref}

\baselineskip 10pt

\end{document}